\documentclass[10 pt, conference]{ieeeconf}
\IEEEoverridecommandlockouts      
\usepackage{amsmath,amssymb,amsfonts}
\usepackage{tikz}
\usepackage{verbatim}
\usepackage[all]{xy}
\usepackage{multirow}
\usepackage{paralist}
\usepackage{wrapfig}
\usepackage{array}
\usepackage{booktabs}

\usepackage{url}

\usepackage{wasysym}
\usepackage{hyperref}
\usepackage{xcolor}
\usepackage{etex}

\newtheorem{remark}{Remark}

\usetikzlibrary{fit}
\usetikzlibrary{backgrounds}
\usetikzlibrary{shapes}
\usetikzlibrary{matrix}
\usetikzlibrary{patterns}
\usetikzlibrary{calc,arrows,automata}
\tikzstyle{State}=[circle, thick, minimum size=0.6cm, inner sep=0cm,draw=black]
\tikzstyle{BState}=[circle, thick, minimum size=0.8cm, inner sep=0cm,draw=black]
\tikzstyle{RState}=[circle, very thick, minimum size=0.8cm, inner sep=0cm,draw=red]

\renewcommand{\baselinestretch}{0.96}

\newcommand{\Supp}{\mathrm{Supp}}

\newcommand{\prior}{D}

\newcommand{\game}{G}
\newcommand{\Last}{\mathsf{Last}}
\newcommand{\plays}{\Omega}

\newcommand{\states}{\mathcal{S}}
\newcommand{\Act}{\mathcal{A}}
\newcommand{\trans}{\mathcal{T}}
\newcommand{\Obs}{\mathcal{Z}}
\newcommand{\obsmap}{\mathcal{O}}
\newcommand{\distr}{\mathcal{D}}

\newcommand{\straa}{\sigma}
\newcommand{\powset}{\mathcal{P}}

\newcommand{\BoolRec}{\mathsf{BoolRec}}

\newcommand{\SetRec}{\mathsf{SetRec}}

\newcommand{\set}[1]{\{#1\}}

\newcommand{\wh}{\widehat}
\newcommand{\nat}{\mathbb{N}}

\newcommand{\goodObs}{\mathit{Good}}
\newcommand{\badObs}{\mathit{Bad}}
\newcommand{\pos}{\mathit{Position}}
\newcommand{\rockType}{\mathit{RockType}}

\title{Qualitative Analysis of POMDPs with Temporal Logic Specifications for Robotics Applications$^{1}$}
\author{Krishnendu Chatterjee$^{2}$ \and Martin Chmel\'{i}k$^{2}$ \and Raghav Gupta$^{3}$ \and Ayush Kanodia$^{3}$
\thanks{$^{1}$ The research was supported by Austrian Science Fund (FWF) Grant No P 23499-N23, FWF NFN Grant No S11407-N23 (RiSE), and ERC Start grant (279307: Graph Games).}
\thanks{$^{2}$ IST Austria (Institute of Science and Technology Austria), Klosterneuburg, Austria}
\thanks{$^{3}$ Indian Institute of Technology, Bombay, India}
}

\begin{document}
\maketitle
\thispagestyle{plain}
\pagestyle{plain}


\begin{abstract}
We consider partially observable Markov decision processes (POMDPs), that 
are a standard framework for robotics applications to model uncertainties
present in the real world, with temporal logic specifications.
All temporal logic specifications in linear-time temporal logic (LTL) can be 
expressed as parity objectives.
We study the qualitative analysis problem for POMDPs with parity objectives 
that asks whether there is a controller (policy) to ensure that the objective 
holds with probability~$1$ (almost-surely).
While the qualitative analysis of POMDPs with parity objectives is undecidable,
recent results show that when restricted to finite-memory policies the problem 
is EXPTIME-complete. 
While the problem is intractable in theory, we present a practical approach to 
solve the qualitative analysis problem.
We designed several heuristics to deal with the exponential complexity, 
and have used our implementation on a number of well-known POMDP examples for 
robotics applications.
Our results provide the first practical approach to solve the qualitative 
analysis of robot motion planning with LTL properties in the presence of uncertainty.
\end{abstract}

\section{Introduction}

\noindent{\bf POMDPs and robotics tasks.}
\emph{Discrete-time Markov decision processes} (MDPs) are standard models for probabilistic 
systems with both probabilistic and nondeterministic behavior~\cite{Howard,FV97}:
nondeterminism represents the freedom of the controller (such as controller for robot
motion planning) to choose a control action, while the probabilistic component of the 
behavior describes the response to control actions. 
In \emph{discrete-time partially observable MDPs} (POMDPs) the state space is 
partitioned according to observations that the controller can observe,
 i.e., given the current state, the controller can only view the observation 
of the state (the partition the state belongs to), but not the precise 
state~\cite{PT87}.
Accounting for uncertainty is a challenging problem for robot motion planning~\cite{probabilisticBook},
and POMDPs provide the appropriate mathematical framework to model a wide variety of 
problems in the presence of uncertainty, including several complex robotics tasks such 
as grasping~\cite{HKL07}, navigation~\cite{RBFT99}, and exploration~\cite{SS05}.
The analysis of POMDPs has traditionally focused on finite-horizon objectives~\cite{PT87} 
(where the problem is PSPACE-complete) or discounted reward objectives~\cite{S04,KHL08}.
While the analysis problem for POMDPs is intractable in theory, and was only applicable
to relatively small problems, a practical approach for POMDPs with discounted reward and 
finite-horizon objectives that scales to interesting applications in robotics was 
considered in~\cite{GMK13}.

\smallskip\noindent{\bf Temporal logic properties.}
While finite-horizon and discounted reward objectives represent an important
class of stochastic optimization problems, several problems in robotics require
a different form of specification, namely, \emph{temporal logic specifications}.
In a temporal logic specification, the objective (or the goal for the control) is 
specified in terms of a \emph{linear-time temporal logic} (LTL) formula that expresses 
the desired set of paths in the POMDP.
While the applicability of temporal logic in robotics was advocated already in~\cite{AM95},
more concretely it was shown in~\cite{FKP05} that LTL provides the mathematical
framework to express properties such as motion sequencing, synchronization, and 
temporal ordering of different motions.
The analysis of (perfect-observation) continuous time systems (such as hybrid systems) with 
temporal logic specifications for robotics tasks have been considered in several works~\cite{FKP05_2,HHW97}.

\smallskip\noindent{\bf POMDPs with parity objectives: Analysis problems.}
POMDPs with discounted reward (or finite-horizon) objectives do not 
provide the framework to express properties like temporal ordering of events 
(which is conveniently expressed in the temporal logic framework). 
On the other hand, perfect-observation continuous time systems do not provide 
the appropriate framework to model uncertainties (in contrast uncertainties are 
naturally modeled as partial observation in POMDPs).
Thus POMDPs with temporal logic specifications expressed in LTL is a very relevant 
and general framework for robotics applications which we consider in this work.
Every LTL formula can be converted into a deterministic parity automaton~\cite{S88},
and hence we focus on POMDPs with parity objectives.
In a parity objective, every state of the POMDP is labeled by a non-negative
integer priority and the goal is to ensure that the minimum priority visited
infinitely often is even.
The analysis problem of POMDPs with parity objectives can be classified as 
follows:
(1)~the \emph{qualitative analysis} asks whether the objective can be ensured with 
probability~1 (\emph{almost-sure satisfaction});
and (2)~the \emph{quantitative analysis} asks whether the objective can be ensured
with probability at least $\lambda \in (0,1)$.
The qualitative analysis is especially important for the following reasons:
first, since probability~1 satisfaction of an objective is the strongest form of 
satisfaction,  almost-sure satisfaction provides the strongest guarantee to satisfy the 
objective; and second, the qualitative analysis is robust with respect to modeling errors 
in the transition probabilities.  
For details of significance and importance of the qualitative analysis problem for MDPs 
and POMDPS see~\cite{CCT13,KHJS13} (also see Remark~\ref{rem:sig} in Appendix).

\smallskip\noindent{\bf Previous results.}
It follows from~\cite{BGB12} that the qualitative-analysis problem is 
undecidable for POMDPs with parity objectives. 
However, recently in~\cite{CCT13} it was shown that when restricted
to finite-state controllers, the qualitative-analysis problem for POMDPs
with parity objectives is EXPTIME-complete.
In most practical applications, the controller must be a \emph{finite-state} 
one to be implementable. 
Thus for all practical purposes the relevant question is the existence of finite-state 
controllers.
However, the quantitative analysis problem for POMDPs with parity objectives is 
undecidable even when restricted to finite-state controllers~\cite{Pazbook,CCT13}.

\smallskip\noindent{\bf Our contributions.}
In this work we present a practical approach to solve POMDPs with parity objectives, 
that given a POMDP and a parity objective, decides whether there exists a finite-state 
controller that ensures almost-sure winning satisfaction. 
If such a controller exists, our algorithm outputs a witness controller.   
While the problem we consider is EXPTIME-complete~\cite{CCT13} and hence intractable in theory, 
we developed a number of heuristics (practical approaches) over the exponential-time algorithm 
proposed in~\cite{CCT13}. 
Our heuristics enabled us to deal with the exponential complexity of several practical 
examples relevant to robotics applications.
We implemented our approach and ran our implementation on a number of POMDPs collected 
throughout the literature with temporal logic properties to express classical specifications 
required for robot motion planning. 
Our results show that all the examples can be solved quite 
efficiently, and our implementation could solve the representative large POMDP 
examples of~\cite{GMK13,LCK95} with the classical temporal logic 
specifications for robotics applications.

\smallskip\noindent{\bf Related work.}
POMDPs with discounted reward (or finite-horizon) objectives~\cite{KHL08,S04} have been studied 
deeply in the literature and also applied in robotics tasks~\cite{GMK13,KGFP09,kaelbling1998planning}.
On the other hand, analysis of continuous time stochastic systems with temporal logic 
properties for robotics applications have also been considered~\cite{FKP05_2,HHW97}.
The works of~\cite{FKP05,GJD13,CB13} consider partial-observation models, but not POMDPs, for robotics 
tasks.
However, the general model of POMDPs with temporal logic properties for robotics tasks 
was not considered before, and we provide the first practical approach for qualitative analysis 
of POMDPs with temporal logic properties.



\newcommand{\B}{\mathcal{B}}

\newcommand{\eventually}{\lozenge\,}
\newcommand{\always}{\square\,}
\newcommand{\nextltl}{\bigcirc\,}
\newcommand{\until}{\mathcal{U}\,}

\section{Definitions}
Given a finite set $X$, we denote by $\powset(X)$ the set of subsets of $X$,
i.e., $\powset(X)$ is the power set of $X$.
A probability distribution $f$ on a finite set $X$ is a function $f:X \to [0,1]$ such 
that $\sum_{x\in X} f(x)=1$, and we denote by  $\distr(X)$ the set of 
all probability distributions on $X$. For $f \in \distr(X)$ we denote by $\Supp(f)=\set{x\in X \mid f(x)>0}$
the support of~$f$.

\smallskip\noindent{\bf POMDPs.}
A discrete-time \emph{partially observable Markov decision process (POMDP)} is modeled as a tuple
$\game=(\states,\Act,\trans,\Obs,\obsmap,s_0)$ where: (i)~$\states$ is a finite set of states; 
 (ii)~$\Act$ is a finite alphabet of \emph{actions};
 (iii)~$\trans: \states \times\Act \rightarrow \distr(\states)$ is a 
 \emph{probabilistic transition function} that given a state~$s$ and an
 action $a \in \Act$ gives the probability distribution over the successor 
 states, i.e., $\trans(s,a)(s')$ denotes the transition probability from state
 $s$ to state $s'$ given action $a$; 
 (iv)~$\Obs$ is a finite set of \emph{observations}; 
 (v)~$\obsmap:S\rightarrow \Obs$ is a deterministic \emph{observation function} that 
  maps every state to an observation; and 
 (vi)~$s_0$ is the initial state. 
For more general types of the observation function see Remark~\ref{rem:obs} in the Appendix.

\smallskip\noindent{\bf Plays and belief-supports.}
A \emph{play} in a POMDP is an infinite sequence $(s_0,a_0,s_1,a_1,\ldots)$ such that for all $i\geq 0$
we have $\trans(s_i,a_i)(s_{i+1})>0$. We write $\plays$ for the set of all plays. 
For a finite prefix $w \in (\states \cdot \Act)^* \cdot \states$ of a play,
we denote by $\Last(w)$ the last state of $w$.
For a finite prefix $w=(s_0,a_0,s_1,a_1,\ldots,s_n)$ 
we denote by 
$\obsmap(w)=(\obsmap(s_0),a_0,\obsmap(s_1),a_1,\ldots,\obsmap(s_n))$ 
the observation and action sequence associated with $w$.
For a finite sequence $\rho=(z_0,a_0,z_1,a_1,\ldots,z_n)$ of observations and actions, the \emph{belief-support} 
$\B(\rho)$ 
after the prefix $\rho$ is the set of states in which a finite prefix 
of a play is with positive probability after the sequence $\rho$ of observations and actions, 
i.e., $\B(\rho)=\set{s_n=\Last(w) \mid w=(s_0,a_0,s_1,a_1,\ldots,s_n), w \text{
is a prefix of a play, and for all } 0\leq i \leq n. \; \obsmap(s_i)=z_i}$. 

\smallskip\noindent{\bf Policies.}
A \emph{policy} is a recipe to extend prefixes of plays and 
is a function $\sigma: (S\cdot A)^* \cdot S \to \distr(A)$ that given a finite 
history (i.e., a finite prefix of a play) selects a probability distribution 
over the actions.
Since we consider POMDPs, policies are \emph{observation-based}, i.e., 
for all histories $w=(s_0,a_0,s_1,a_1,\ldots,a_{n-1},s_n)$ and 
$w'=(s_0',a_0,s_1',a_1,\ldots,a_{n-1},s_n')$ such that for all 
$0\leq i \leq n$ we have $\obsmap(s_i)=\obsmap(s_i')$ (i.e., $\obsmap(w) = \obsmap(w')$), we must have 
$\sigma(w)=\sigma(w')$.
In other words, if the observation sequence is the same, then the policy 
cannot distinguish between the prefixes and must play the same. 
We now present an equivalent definition of observation-based policies  
such that the memory of the policy is explicitly specified, and 
will be required to present finite-memory policies.

\smallskip\noindent{\bf Policies with memory and finite-memory policies}
A \emph{policy} with memory is a tuple $\sigma=(\sigma_u,\sigma_n,M,m_0)$ where:
 (i)~\emph{(Memory set).} $M$ is a denumerable set (finite or infinite) of memory elements (or memory states).
 (ii)~\emph{(Action selection function).} The function $\sigma_n:M\rightarrow \distr(\Act)$ is the 
	\emph{next action selection function} that given the current memory 
	state gives the probability distribution over actions.
 (iii)~\emph{(Memory update function).} The function $\sigma_u:M\times\Obs\times\Act\rightarrow \distr(M)$ 
 is the \emph{memory update function} that given the current memory state, 
 the current observation and action, updates the memory state probabilistically.
 (iv)~\emph{(Initial memory).} The memory state $m_0\in M$ is the initial memory state.
A policy is a \emph{finite-memory} policy if the set $M$ of memory elements is finite. A policy
is \emph{memoryless} if the set of memory elements $M$ contains a single memory element.

\smallskip\noindent{\bf Objectives.}
An objective specifies the desired set of paths (or behaviors) in a POMDP.
A common approach to specify objectives is using LTL formulas~\cite{CGP99},
and LTL formulas can express all commonly used specifications in practice. 
We first give some informal examples of objectives used in the literature~\cite{FKP05,RPK13,CTB12},
and we use the following notation for LTL temporal operators such as eventually ($\eventually$), always ($\always$), next ($\nextltl$) and until ($\until$).

\begin{compactitem}
\item \textbf{Liveness objective:} Given a set of goal states $T \subseteq \states$, the liveness 
objective is to reach the goal states (in LTL notation $\eventually T$).

\item \textbf{Safety:} Given a set of safe states $L \subseteq \states$, the objective is to stay in the safe states
(in LTL notation $\always L$). 

\item \textbf{Reach a goal while avoiding obstacles:} The objective generalizes the previous two objectives and is defined by
 a set of obstacles $O_1, O_2, \ldots, O_n$, where every
obstacle $O_i$ for $1 \leq i \leq n$ is defined by a set of states  $O_i \subseteq \states$, and a 
set of goal states $T \subseteq \states$, the objective is to reach the goal states while avoiding all the obstacles
(in LTL notation $\neg ( O_1 \vee O_2 \ldots \vee O_n) \until T$). 

\item \textbf{Sequencing and Coverage:} Given a sequence of locations $S_1, S_2, \ldots, S_n$ where every
location $S_i$ for $1 \leq i \leq n$ is given by a set of states $S_i \subseteq \states$, the \emph{sequencing} objective is
to visit all the locations in the given order (in LTL notation $\eventually ( S_1 \wedge \eventually(S_2 \wedge \eventually (\ldots (\eventually S_n )))$).
The \emph{coverage} objective is
to visit all the locations in any order (in LTL notation $\eventually  S_1 \wedge \eventually S_2 \wedge \ldots \wedge\eventually S_n $).

\item \textbf{Recurrence:} Given a set of states $S \subseteq \states$, the objective is to visit the set of states
$S$ infinitely often (in LTL notation $\always \eventually S$).
\end{compactitem}

\smallskip\noindent{\bf Parity objectives.}
We will focus on POMDPs with parity objectives, since every LTL formula 
can be translated to a deterministic parity automaton~\cite{S88,P06}.
Given a POMDP, an LTL formula, and an equivalent deterministic parity automaton 
for the formula, the synchronous product of the POMDP and the automaton gives us a 
POMDP with a parity objective.
For all the above objectives mentioned, the translation to parity objectives is 
simple and straightforward.
\begin{compactitem}
\item\textbf{Parity objectives:}
Given a priority function 
$p :\states \rightarrow \nat$ assigning every state a non-negative priority.
A play $\rho = (s_0,a_0,s_1,a_1,s_2,a_2,\ldots)$ is winning (i.e., satisfies the parity objective) if the minimum priority 
appearing infinitely often in the play is even. 
\item\textbf{coB\"uchi objectives:} The coB\"uchi objectives are a special case of parity objectives, where the priority function assigns only values $1$ and $2$.
\end{compactitem}
We consider the special case of coB\"uchi objectives because our algorithmic analysis will reduce
POMDPs with parity objectives to POMDPs with coB\"uchi objectives.

\smallskip\noindent{\bf Qualitative analysis.}
Given a policy $\sigma$, let $\mathbb{P}_{s_0}^{\sigma}(\cdot)$ denote the probability measure obtained 
by fixing the policy in the POMDP~\cite{Var85}.
A policy $\straa$ is \emph{almost-sure winning} for a parity objective $\varphi$ 
if $\mathbb{P}_{s_0}^{\sigma}(\varphi) =1$.
The \emph{qualitative} analysis problem given a POMDP and a parity objective asks for the 
existence of an almost-sure winning policy. 
For significance of qualitative analysis see Remark~\ref{rem:sig} in Appendix.

\section{Existing Results}
We first summarize the existing results.

\smallskip\noindent{\bf Previous results.}
The qualitative analysis of POMDPs with parity objectives is undecidable~\cite{BGB12};
and the problem is EXPTIME-complete when restricted to the practical case
of finite-memory policies~\cite{CCT13}.
It was also shown in~\cite{CCT13} that the traditional approach of subset construction 
does not provide an algorithmic solution for the problem.
The quantitative analysis problem of POMDPs with parity objectives is 
undecidable even for finite-memory policies~\cite{Pazbook,CCT13}.

\smallskip\noindent{\bf Algorithm from~\cite{CCT13}.}
We now summarize the key ideas of the algorithm for qualitative analysis with finite-memory 
policies presented in \cite{CCT13}.

\smallskip\noindent{\bf Step 1: Reduction to coB\"uchi objectives.}
The results of~\cite{CCT13} present a polynomial-time reduction from 
POMDPs with parity objectives to POMDPs with coB\"uchi objectives for qualitative analysis 
under finite-memory policies. 

\smallskip\noindent\textbf{Step 2: Solving POMDPs with coB\"uchi objectives.}
The main algorithmic result of~\cite{CCT13} is solving the qualitative analysis problem 
for POMDPs with coB\"uchi objectives.
The key proof shows that if there exists a finite-memory almost-sure winning policy, 
then there exists a \emph{projected} policy $\sigma=(\sigma_u,\sigma_n,M,m_0)$ that is also almost-sure winning and the 
projected policy requires at most $2^{6 \cdot \vert \states \vert}$ memory states, i.e., $M \leq 2^{6 \cdot \vert \states \vert}$.
 The fact that given a POMDP there exists a bound on the number of memory elements required by an almost-sure
winning policy already establishes the decidability result. Another consequence of the result is
that \emph{projected} policies are sufficient for almost-sure winning in POMDPs. The knowledge of the
memory elements, the structure of memory-update function $\sigma_u$, and the action-selection function $\sigma_n$
are crucial for the last step of the algorithm and our results.
The key components of the projected policy memory elements $M$ are as follows:
(i)~The first component is the \emph{belief-support}, i.e., the subset of states in which the POMDP is with positive probability.
(ii)~The second component (namely $\BoolRec$) denotes whether a state and the current memory is recurrent or not, i.e., 
if reached, will be almost-surely visited infinitely often.
(iii)~Finally, the third component (namely $\SetRec$) stores a mapping from the states of the POMDP to the 
priority set of the reachable recurrent classes.
The memory elements $m \in M$ will be written as follows: $m = (Y,B,L)$, where $Y \subseteq \states$ is the belief-support 
component, $B: \states \rightarrow \{0,1\}$ is the $\BoolRec$ component, and the $\SetRec$ component is $L: \states \rightarrow \powset(\powset(\prior))$,
 where $\prior \subseteq \nat$ is the set of priorities used by the parity objective, in particular for coB\"uchi objectives
 we have $\prior = \{1,2\}$.

\smallskip\noindent{\bf Step 3: Solving synchronized product.}
It follows from the previous steps that the qualitative analysis of POMDPs reduces to the problem of deciding
whether in a given POMDP~$\game$ with a coB\"uchi objective exists a \emph{projected} almost-sure winning policy.
The final algorithmic idea is to construct an exponential POMDP, called \emph{belief-observation} POMDP~$\wh{\game}$, which 
intuitively is a synchronized product of the original POMDP and the most general projected policy. Intuitively, the advantage of
considering the synchronized product is that the memory elements of the \emph{projected} policy are already
present in the state space of the POMDP $\wh{\game}$. It follows that if there exists a \emph{projected} almost-sure winning policy
 in the POMDP, then there
exists a memoryless almost-sure winning policy in the synchronized product POMDP, and vice versa.
Finally, to decide whether there exists a memoryless almost-sure winning policy in the  belief-observation 
POMDP $\wh{\game}$ can be solved in polynomial time.



\section{Practical Approaches and Heuristics}
\label{sec:tool}
In this section we present the key ideas and heuristics that allowed efficient 
implementation of the algorithmic ideas of~\cite{CCT13}.
Step~1 and Step~3 of the algorithmic ideas of~\cite{CCT13} are polynomial time 
and we have implemented the algorithms proposed in~\cite{CCT13}.
Step~2 of the algorithmic ideas of~\cite{CCT13} is exponential and posed the main 
challenge for efficient implementation. 
We employed several heuristics to make Step~2 practical and we describe them below.

\smallskip\noindent{\bf Heuristics.} Our heuristics are based on ideas to reduce
the number of memory elements $M$ required by the  \emph{projected} policy. As the projected
 policy plays in a structured way,
we exploit the structure to reduce its size employing the following heuristics.

\begin{compactenum}
\item The first heuristic reduces the size of the memory set $M$ of the \emph{projected} policy.
Intuitively, instead of storing the  mappings $\BoolRec$ and $\SetRec$ for every state $s \in \states$ of the POMDP,
we store the mappings restricted to the current belief-support, i.e., given a memory element $m = (Y,B,L)$ we consider
the $\BoolRec$ component $B$ to be of type $B: Y \rightarrow \{0,1\}$, similarly for the $\SetRec$ component
$L$ we restrict the domain of the function to $Y$, i.e., we have $L: Y \rightarrow \powset(\powset(\prior))$ ($\prior$
denotes the set of priorities).
Intuitively, for all states that are not in the belief-support $Y$, the probability of being in them is~$0$. 
Therefore, the information stored about these states is not relevant for the \emph{projected} policy.
The size of the current belief-support is often significantly smaller than the number of states,
 as the size of the belief-support 
is bounded by the size of the largest observation in the POMDP, i.e., the size of the belief-support is bounded 
by $\max_{z\in \Obs} \vert \obsmap^{-1}(z)\vert$.  
It follows, that it also improves the theoretical bound on the size of the belief-observation
 POMDP presented in \cite{CCT13}.

\item The second reduction in memory relies on the following intuition: given a memory element $m=(Y,B,L)$ 
by the first heuristic we store the mappings 
only for the states of the current belief-support $Y$, and the belief-support represents exactly the states that the POMDP is 
in with positive probability. An important property of the \emph{projected} policy is that
the $\SetRec$ function $L: S \rightarrow \powset(\powset(\prior))$ corresponds to the priority set of reachable recurrent classes. 
Intuitively, for every state $s \in Y$,
we have that every reachable recurrent class of the \emph{projected} policy from state $s$ and memory $m$ will have the priority set
of its states $U$ in $L(s)$, and for every priority set in $U \in L(s)$, there exists a recurrent class with a priority set $U$, 
that is reachable with positive probability. Therefore, all the reachable recurrent classes according to the
$\SetRec$ mapping are reached with positive probability with the projected policy. 
As the projected policy is almost-sure winning it follows that all the reachable recurrent classes 
must also be winning.
Since the objective in the POMDP~$\wh{\game}$ is a coB\"uchi objective, we have that a winning recurrent class must consist 
 only of coB\"uchi states (only states with priority $2$). 
Therefore, we can restrict the the range of the $\SetRec$ mapping to a singleton $\{\{2\}\}$. 
It follows that we do not have to consider the $\SetRec$ component of the \emph{projected} policy at all.
\end{compactenum}
The main contribution of the above two ideas is that the running time is no longer exponential in the 
number of the states of the POMDP, but rather in the largest belief-support reachable.
Since in many practical cases, the largest belief-support reachable is quite small, our heuristics on top of 
the algorithmic ideas of~\cite{CCT13} provide an efficient solution for several examples 
(as illustrated in Section~\ref{sec:results}).


\section{Case Studies}
\label{sec:results}
We implemented all the algorithmic ideas of~\cite{CCT13} along with the improvements
as described in Section~\ref{sec:tool}.
Our implementation is in Java, and we have 
tested it on a number of well-known examples
from the literature. The computer we used is equipped with 8GB of memory and
a quad-core i7 2.0 GHz CPU.
Detailed descriptions of all our examples are provided in the Appendix 
(we present succinct descriptions below).

\smallskip\noindent\textbf{Space Shuttle.}
The space shuttle example was originally introduced in~\cite{C92}, and it models a simple space 
shuttle that delivers supplies to two space stations. 
There are three actions that can be chosen: \emph{go-forward, turn-around, and backup}.
The goal is to visit the two stations delivering goods infinitely often and avoid bumps 
(trying to go forward when facing a station). 
The docking is simulated by backing up into the station.
The parity objective has~3 priorities and is as follows: 
traveling through the space has priority~3, 
delivering goods to the station that was not visited has priority~2, and 
bumping has priority 1. Therefore, the objective is to control the shuttle in a way 
that it delivers supplies to both stations infinitely often, while bumping into the space 
station only finitely often.
The POMDP corresponding to the one introduced in~\cite{C92} has $11$ states. 
Along with the original POMDP of~\cite{C92} we also consider two variants 
with $13$ and $15$ states, respectively, that intuitively increases the distance to travel between the stations, 
and this affects the amount of uncertainty in the system and leads to larger
belief-support sets (and hence longer running times).
The POMDPs after the coB\"uchi reduction have $23$, $27$, and $31$ states, respectively, and were solved
in~$0.07$, $0.24$, and~$1.01$ seconds, respectively.


\smallskip\noindent\textbf{Cheese Maze} \cite{LCK95}.
The problem is given by a maze modeled as a POMDP. The movement in the POMDP is deterministic in all 
four directions -- north, south, east, and west. Movements that attempt to move outside
of the maze have no effect on the position. The observations correspond
to what would be seen in all four directions immediately adjacent to the location. Some of the states
are marked as \emph{goal} states and some are marked as \emph{bad} states. Whenever a 
goal state is visited the game is restarted. The objective is to visit the goal states infinitely 
often while the bad states should be visited only finitely often. The original maze introduced in~\cite{LCK95} has 
$11$ states. 
We also consider extensions of the maze POMDP that has $23$ states. 
Depending on the amount of uncertainty about the current position after 
restarting the game we have three variants, namely, easy, medium, and hard, 
for both sizes of the maze.
The number of states the POMDPs have after the coB\"uchi reduction is $23$ and $47$ states,
respectively and all the cases were solved in less than $7$ seconds.

\smallskip\noindent\textbf{Grid.}
The example is based on a problem introduced in~\cite{LCK95} and consists
of a grid of locations. As in the previous example some of the locations
are \emph{goal} locations and some are marked as \emph{bad} locations. Whenever a
goal location is reached the game is restarted to the initial state. The objective is to visit the goal locations
infinitely often while visiting the bad locations only finitely often. In the very first step
the placement of the goal and bad states is done probabilistically and does not change during the play. The goal is to
learn the maze while being partially informed about its surroundings. We consider five
variants that differ in size, i.e., the grid $4 \times 4$ is has 33 states,
$5 \times 5$ has 51 states, $6\times6$ has $73$ states, $7\times7$ has 99 states, and $8\times8$ has $129$ states. 
After the coB\"uchi reduction the POMDPs have~$67$, $103$, $147$, $199$, and $259$ states, respectively. All the variants were solved in less than $6$ seconds.

\smallskip\noindent\textbf{RockSample problems.}
We consider a modification of the RockSample problem introduced in~\cite{SS04} and used later in~\cite{BG09}. It is a scalable problem that models
 rover science exploration. The rover is equipped with a limited amount of fuel and can
 increase the amount of fuel only by sampling rocks in the immediate area.
   The positions of the rover and the rocks are known, but only some of the rocks can increase the amount of fuel; we will call these 
   rocks good. The type of the rock is not known to the rover, until the rock is sampled.
Once a good rock is used to increase the amount of fuel, it becomes temporarily a bad rock until all other good rocks are sampled. We consider
   variants with different maximum capacity of the rover's fuel tank.
An instance of the RockSample problem is parametrized with two parameters $[n,k]$: map size $n \times n$ and $k$ rocks 
is described as RockSample[n,k]. The POMDP model of RockSample[n,k] is as follows:
The state space is the cross product of $2k+1+c$ features: 
$\pos = \{(1, 1), (1, 2), . . . , (n, n)\}$, $2*k$ binary features $\rockType_i = \{\goodObs, \badObs\}$ that indicate which of the rocks are good and which rocks
are temporarily not able to increase the amount of fuel, and $c$ is the amount of fuel remaining in the fuel tank.
There are four observations: the unique observation for the initial state, two observations
to denote whether the rock that is sampled is good or bad, and the last observation is for all the remaining states.
After the coB\"uchi reduction the POMDPs have~$1025$, $1281$, $3137$, and $3921$ states, respectively.
 All the variants were solved in less than $15$ seconds.

\smallskip\noindent\textbf{Hallway problems.}
We consider two versions of the Hallway problems introduced in~\cite{LCK95} and used later in~\cite{S04,SS04,BG09}.
The idea behind both of the Hallway problems, is that there is an agent 
wandering around an office building. 
It is assumed that the locations have been discretized so that there are a finite number of 
locations where the agent could be. The agent has a small finite set of actions it can take, 
but these only succeed with some probability. Additionally, the agent is equipped 
with very short range sensors to provide it only with information about whether it is 
adjacent to a wall. The sensors can ''see'' in four directions: forward, left, right,  and backward.
 Note that these observations are relative to the 
current orientation of the agent (N, E, S, W). In these problems the location in the building and the agent's current orientation comprise the states.
There are four dedicated areas in the office, denoted by letters $A$, $B$, $C$, and $D$. We consider four objectives in both the 
Hallway problems:
\begin{compactitem}
\item \emph{Liveness:} requires that the $D$-labeled area is reached.
\item \emph{Sequencing and avoiding obstacles:} requires that first the $A$-labeled area is visited, followed by the $B$-labeled area
and finally the $D$-labeled area is visited while avoiding the $C$-labeled area. 
\item \emph{Coverage:} requires that the $A$, $B$, and $C$-labeled areas are all visited in any order.
\item \emph{Recurrence:} requires that both the $A$ and $C$-labeled areas are visited infinitely often.
\item \emph{Recurrence and avoidance:} requires that both $A$ and $D$-labeled areas are visited infinitely often, while visiting $B$ and $C$-labeled states
only finitely many times.
\end{compactitem}
The size of the POMDPs for the smaller Hallway problem depends on the objective and has up to 453 states. In the Hallway~2 problem the POMDPs
have up to $709$ states. All the variants were solved in less than $21$ seconds.


\smallskip\noindent\textbf{Maze navigation problems.}
We consider three variants of the mazes introduced in~\cite{GMK13}. Intuitively, the robot navigates itself in a grid
discretization of a 2D world. The robot can choose from four noise free actions north, east, south, and west. In every maze
there are $4$ highlighted areas that are labeled with letters $A$, $B$, $C$, and $D$. The objectives for the robot are the same
 as in the case of the Hallway problems.
The state space of the problem consists of the possible grid locations times the number of states of the parity automaton that specifies
the objective. The robot moves from the unique initial states uniformly at random under all actions to all the locations labeled with "+".
Beside the highlighted areas, the robot does not receive any feedback from the maze.
In locations where the robot attempts to move outside of the maze or in the wall, the position of the robot remains unchanged.
We consider the same objectives as in the case of the Hallway problems.
The sizes of the models go up to $641$ states, and all the variants were solved in less than $12$ minutes.

\begin{table}[h!]

\centering
\resizebox{\linewidth}{!}{
\begin{tabular}{|r|c|c|c|}

\toprule
& Time & $\vert \states \vert$ &  \begin{tabular}[x]{@{}c@{}} $\vert \states \vert$ after \\ the reduction\end{tabular}\\

\midrule
\multicolumn{1}{|l|}{\textbf{Space Shuttle:} (3 act., 7 obs.)}  & & & \\
small & 0.07s & 11  & 23\\
medium & 0.24s & 13  & 27 \\
large & 1.01s & 15  & 31\\
\midrule
\multicolumn{1}{|l|}{\textbf{Small Cheese maze:} (4 act., 7 obs.)}  & & & \\
easy & 0.03s & 11  & 23\\
medium & 0.05s & 11  & 23 \\
hard & 0.07s & 11  & 23\\
\midrule
\multicolumn{1}{|l|}{\textbf{Large Cheese maze:} (4 act., 7 obs.)}  & & & \\
easy & 0.22s & 23  & 47\\
medium & 0.72s & 23  & 47 \\
hard & 6.63s & 23  & 47\\

\midrule
\multicolumn{1}{|l|}{\textbf{Grid:} (4 act., 6 obs.)}  & & & \\
$4\times4$ & 0.69s & 33  & 67 \\
$5\times5$ & 1.37s & 51  & 103 \\
$6\times6$ &  2.16s &73   & 147 \\
$7\times7$ &  3.94s &99   &  199\\
$8\times8$ &  5.89s &129   &  259\\
\midrule
\multicolumn{1}{|l|}{\textbf{RS[4,2]:} (4 act., 4 obs.)}  & & & \\
Capacity 3 & 0.46s & 1025  & 1025 \\
Capacity 4 & 1.00s & 1281 & 1281 \\
\midrule
\multicolumn{1}{|l|}{\textbf{RS[4,3]:} (4 act., 4 obs.)}  & & & \\
Capacity 3 & 8.70s & 3137  & 3137 \\
Capacity 4 & 14.98s & 3921 & 3921 \\
\midrule
\multicolumn{1}{|l|}{\textbf{Hallway:} (3 act., 16 obs.)}  & & & \\
Liveness & 0.73s & 120   & 121 \\
Seq. and avoid. obstacles & 1.89s& 276  & 277 \\
Coverage & 2.48s& 453  & 454 \\
Recurrence & 1.18s& 185  & 186 \\
Rec. and avoid. & 5.04s & 180  & 361 \\
\midrule
\multicolumn{1}{|l|}{\textbf{Hallway 2:} (3 act., 19 obs.)}  & & & \\
Liveness & 1.18s & 184  & 185 \\
Seq. and avoid. obstacles & 2.53s & 436  &  437 \\
Coverage & 4.98s& 709  & 710 \\
Recurrence & 1.99s& 281  & 282 \\
Rec. and avoid. & 20.02s & 276  & 553 \\
\midrule
\multicolumn{1}{|l|}{\textbf{Maze A:} (4 act., 6 obs.)} & & & \\
Liveness & 23.03s & 169  & 170 \\
Seq. and avoid. obstacles & 66.24s & 371 & 372 \\
Coverage & 58.50s& 573 & 574 \\
Recurrence & 33.57s& 267 & 268 \\
Rec. and avoid. & 608.25s & 263  & 527 \\
\midrule
\multicolumn{1}{|l|}{\textbf{Maze B:} (4 act., 6 obs.)} & & & \\
Liveness & 16.18s & 154  & 155 \\
Seq. and avoid. obstacles & 103.17s & 380 & 381 \\
Coverage & 84.19s& 641 & 642 \\
Recurrence & 24.68s& 254 & 255 \\
Rec. and avoid. & 668.17s & 258  & 517 \\
\midrule
\multicolumn{1}{|l|}{\textbf{Maze C:} (4 act., 6 obs.)} & & & \\
Liveness & 1.12s & 110  & 111 \\
Seq. and avoid. obstacles & 1.56s & 267 & 268 \\
Coverage & 1.44s& 439 & 440 \\
Recurrence & 0.84s& 200 & 201 \\
Rec. and avoid. & 1.24s & 116  & 233 \\

\bottomrule
\end{tabular}
}
\caption{Obtained Results}
\label{tab:res}
\vspace{-2.5em}
\end{table}

We summarize the obtained results in Table~\ref{tab:res}. For every POMDP we show the running
time of our tool in seconds, the number of states of the POMDP, and finally the number of states
of the POMDP after the reduction to a coB\"uchi objective.

\smallskip\noindent{\em Effectiveness of the heuristics.} 
It follows from~\cite{CCT13} that for solving POMDPs even subset construction is not enough. Hence without the proposed heuristics, at least an exponential subset construction is required (and even more), while explicit subset construction is prohibitive in all our examples. Thus if we turn-off our heuristics, then the implementation does not work at all on the examples. 


\section{Conclusion and Discussion}
In this work we present the first practical approach for qualitative
analysis of POMDPs with temporal logic properties, and show that our 
implementation can handle representative POMDPs that are relevant for 
robotics applications.
A possible direction of future work would be to consider quantitative
analysis: though the quantitative analysis problem is undecidable
in general, an interesting question is to study subclass and design 
heuristics to solve relevant practical cases of quantitative 
analysis of POMDPs with temporal logic properties.

The heuristics we propose exploit the fact that in many case studies where POMDPs 
are used, the uncertainty in the knowledge is quite small (i.e., formally the belief-support
sets are small). While for perfect-information MDPs efficient (polynomial-time) algorithms
are known, our heuristics show that if the belief-support sets are small (i.e., the 
uncertainty in knowledge is small), then even POMDPs with parity objectives with
 a small number of priorities can be solved efficiently. 
The limiting factor of our heuristics is that if the belief-support sets are large,
then due to exponential construction our algorithms will be inefficient in 
practice. 
An interesting direction of future work would be consider methods such as 
abstractions for POMDPs~\cite{FDT14} and combine them with our heuristics to solve
large scale POMDPs with a huge amount of uncertainty.



\bibliographystyle{plain}
\bibliography{diss}
\clearpage


\appendix
\section{Appendix}
\label{chap:app}

We present two remarks, the first remark considers more general types of the observation function in POMDPs. 
The second remark comments on the significance of qualitative analysis.

\begin{remark}[Observations]
\label{rem:obs}
We remark about two other general cases of observations.
\begin{enumerate}
\item \emph{Multiple observations:}
We consider observation function that assigns an observation to 
every state.
In general the observation function $\obsmap: \states \rightarrow 2^{\Obs}  \setminus  \emptyset$ 
may assign multiple observations to a single state.
In that case we consider the set of observations as $\Obs' = 2^{\Obs} \setminus \emptyset$ 
and consider the mapping that assigns to every state an observation 
from $\Obs'$ and reduce to our model.
\item \emph{Probabilistic observations:} 
Given a POMDP $\game=(\states,\Act,\trans,\Obs,\obsmap,s_0)$, another type
of the observation function $\obsmap$ considered in the literature is of type 
$\states \times \Act \rightarrow \distr(\Obs)$, i.e., the state and the action 
gives a probability distribution over the set of observations $\Obs$. 
We show how to transform the POMDP~$\game$ into an equivalent POMDP $\game'$ 
where the observation function is deterministic and defined on states, i.e., of type 
$\states \rightarrow \Obs$ as in our definitions. 
We construct the equivalent POMDP $\game'=(\states',\Act,\trans',\Obs,\obsmap',s_0')$
as follows: 
(i)~the new state space is $\states' = \states \times \Obs$; 
(ii)~the transition function $\trans'$ given a state $(s,z) \in \states'$ and an action $a$ 
is as follows $\trans'((s,z),a)(s',z') = \trans(s,a)(s') \cdot \obsmap(s',a)(z')$; and
(iii) the deterministic observation function for a state $(s,z) \in \states'$ is defined as $\obsmap'((s,z)) 
= z$.
Informally, the probabilistic aspect of the observation function is captured in the 
transition function, and by enlarging the state space by constructing a product with the observations,
we obtain a deterministic observation function only on states.
\end{enumerate}
Thus both the above general cases of observation function can be reduced to observation mapping 
that deterministically assigns an observation to a state, and 
we consider such observation mapping which greatly simplifies the notation.
\end{remark}

\begin{remark}[Significance of qualitative analysis.]
\label{rem:sig}
The qualitative analysis problem is important and significant for the following 
reasons.
First, under finite-memory policies, while the quantitative analysis is undecidable,
the qualitative analysis is decidable.
Second, the qualitative analysis (winning with probability~1) provides the strongest form 
of guarantee to satisfy an objective. 
Finally, the qualitative analysis problem is robust with respect to modeling errors in 
the probability of the transition function.
This is because once a finite-memory policy is fixed, we obtain a Markov chain, and 
the qualitative analysis of Markov chains only depends on the graph structure of 
the Markov chain and not the precise probabilities.
Thus even if the probabilities are not accurately modeled, but the support of the 
transition function does not change, then the solution of qualitative analysis does not 
change either, i.e., the answer of the qualitative analysis is robust with respect to 
modeling errors in precise transition probabilities.
For more details regarding significance of qualitative analysis see~\cite{CCT13,KHJS13}.
\end{remark}

\section{Appendix - Examples}

\subsection{Example - Space Shuttle}

\begin{figure}
  \centering
  \resizebox{\linewidth}{!}{\includegraphics{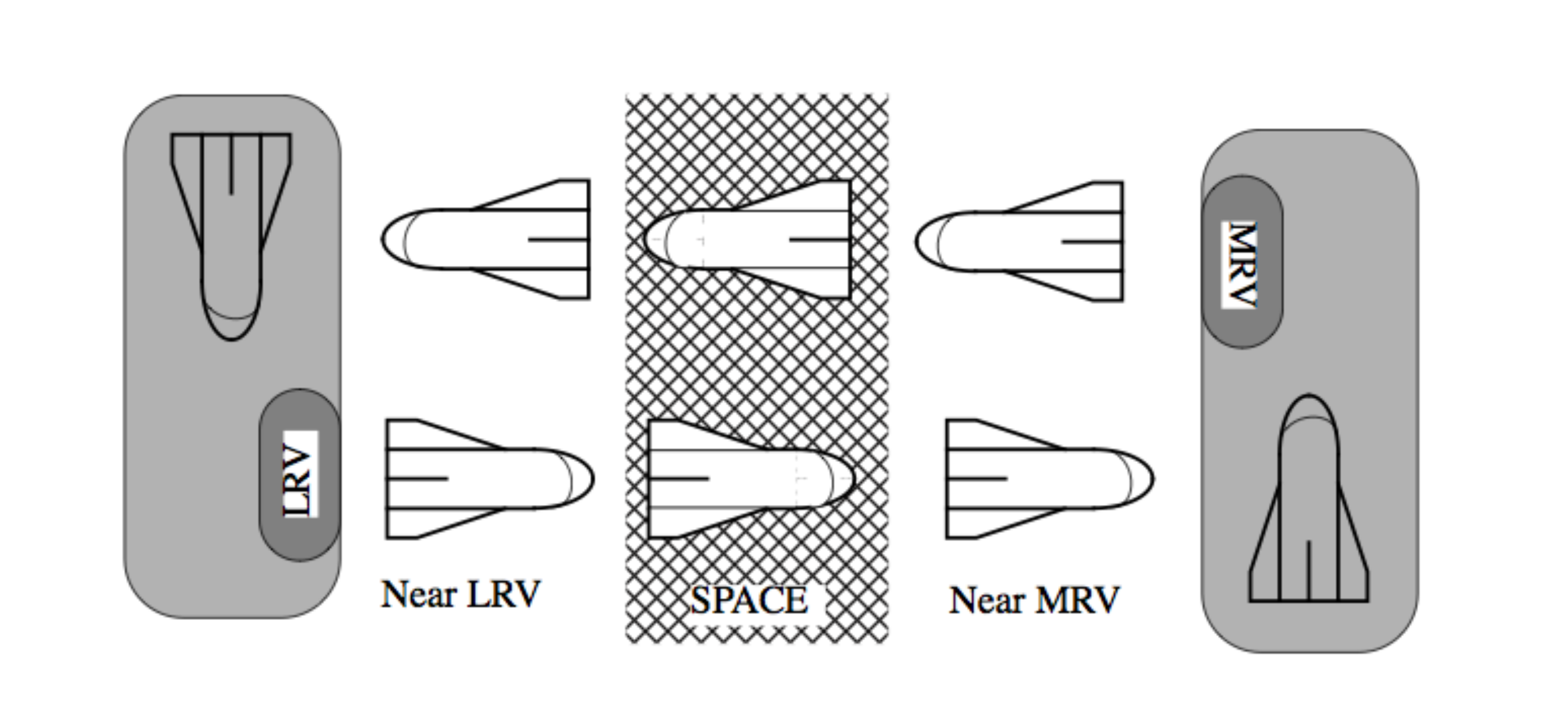}}
  \caption{Space shuttle docking problem}
  \label{fig:shuttle}

\end{figure}

The space shuttle example originally comes from \cite{C92}, and along with the original POMDP we also 
consider slight variants of the model presented in \cite{LCK95}. We will describe the easiest variant
of the problem, the remaining variants will be described in the end of the example.
It models a simple space shuttle docking problem, where 
the  shuttle must dock by backing up into one of the two space 
stations. The goal is to visit both stations infinitely often. Figure~\ref{fig:shuttle} originally comes from \cite{LCK95} and shows a schematic representation of the 
model. The left most and right most states in Figure~\ref{fig:shuttle} are the docking stations, the most recently visited docking station is labeled with MRV, and the least recently visited docking station is labeled with LRV. The property of the model is that whenever a LRV station is visited it automatically changes its state to MRV. Both of the actions go forward and turn around are deterministic.

\smallskip\noindent\textbf{States:}
There are $11$ states in the POMDP, corresponding to the position of the shuttle: 
\verb;0; - docked in LRV;
\verb;1; - just outside space station MRV, front of ship facing station;
\verb;2; - space, facing MRV;
\verb;3; - just outside space station LRV, back of ship facing station;
\verb;4; - just outside space station MRV, back of ship facing station;
\verb;5; - space, facing LRV;
\verb;6; - just outside space station LRV, front of ship facing station;
\verb;7; - docked in MRV, the initial state;
\verb;8; - successful delivery;
\verb;9; - bump into LRV;
\verb;10; - bump into MRV.

\smallskip\noindent\textbf{Observations:}
There are $7$ observations corresponding to what can be seen from the shuttle: \verb;o0; see LRV forward, \verb;o1;, see MRV forward, \verb;o2; docked in MRV, \verb;o3; see nothing, \verb;o4; docked in LRV, \verb;o5; bumping into a docking station, and finally \verb;o6; is observed upon a successful delivery.

\smallskip\noindent\textbf{Actions:} There are three actions that can be chosen: go forward (\verb;f;), turn around (\verb;a;), and backup (\verb;b;). 

\smallskip\noindent\textbf{Transition relation:} 
If the shuttle is facing a station (states \verb;1; and \verb;6;) the backup action succeeds only with probability $0.3$, has no effect with probability $0.4$, and with probability $0.3$ acts like a turn around action. Whenever in space (states \verb;2; and \verb;5;) the backup 
actions succeeds with probability $0.8$, has no effect with probability $0.1$, and with the remaining probability $0.1$ has the same effect
as an combination of turning around and a backup action. Finally, when the shuttle is adjacent to a station and facing away (states \verb;3; and \verb;4;), it has a probability of $0.7$ of actually docking to a station, and with the remaining probability $0.3$ has no effect. In the remaining states \verb'8', \verb'9', and \verb'10' the action effect is deterministic.

\smallskip\noindent\textbf{Objective.} The parity objective of the model is defined by the priority assignment function as follows: all states from \verb;1 ; to \verb;7; 
have priority $3$; the state \verb;8; which represents a successful delivery into a least recently visited station, has priority $2$. Whenever the shuttle is facing a station, it should try to backup into the station as trying to move forward results into a bump into a station represented with states \verb;9; and \verb;10; with priority $1$. Therefore, the objective can be intuitively explained as trying to visit both of the stations infinitely often, while trying to bump only finitely often with probability $1$.

The input file for the POMDP can be downloaded here \url{http://pub.ist.ac.at/pps/examples/Space_shuttle_small.txt}. The more complex variants of this model differ in the number of states that are required to travel through the space, and therefore have higher uncertainty about the position of the shuttle.

\subsection{Example - Cheese Maze}
\begin{figure}
 \centering
  \resizebox{6cm}{!}{\includegraphics{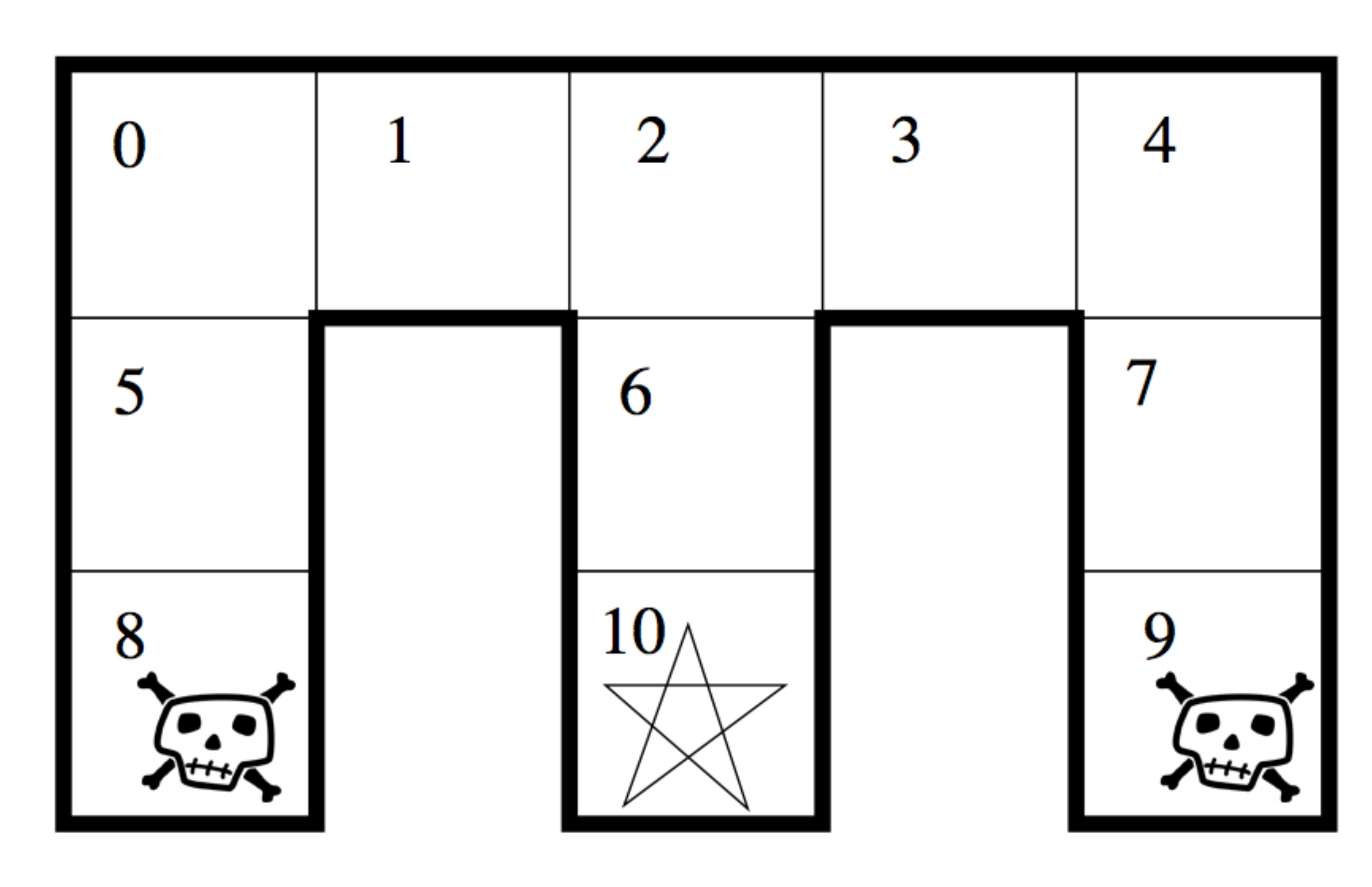}}
  \caption{Cheese maze problem}
  \label{fig:cheese_maze}
\end{figure}

The maze is shown in Figure~\ref{fig:cheese_maze} is introduced in \cite{LCK95}. We will describe only the smallest variant in detail. The goal of the player is to reach a goal state while trying to avoid
poison in bad states. The player is only partially informed, its observation
corresponds to what would be seen in all four directions immediately adjacent  to the location. After the goal state is reached the player
is respawned with positive probability in multiple states of the maze and the game is restarted.

\smallskip\noindent\textbf{States:} There are 
$11$ states in the POMDP that are illustrated on Figure~\ref{fig:cheese_maze}. The game starts in state \verb;6;. The poison is placed in states \verb;8; and \verb;9;, and \verb;10; is the goal state.

\smallskip\noindent\textbf{Observations:}
There are $7$ observations corresponding to what would be seen in all four directions immediately adjacent  to the location, i.e., states \verb;5;, \verb;6;, and \verb;7; do have the same observation. 
The observations are as follows:
\verb;o0;, the walls are NW; 
\verb;o1;, the walls are NS; 
\verb;o2;, the wall is N; 
\verb;o3;, the walls are NE; 
\verb;o4;, the walls are WE; 
\verb;o5;, a poisoned state; and
\verb;o6;, the goal state. 

\smallskip\noindent\textbf{Actions:}
There are four actions available corresponding to the movement
in the four compass directions (north \verb;n;, east \verb;e;, south \verb;s;, west \verb;w;). 

\smallskip\noindent\textbf{Transition relation:}
Actions that attempt to
move outside of the maze have no effect on the position. The rest of the moves is deterministic in all $4$ actions. 

\smallskip\noindent\textbf{Objective:}
The objective is to visit the goal state \verb;10; infinitely often, while getting poisoned in states \verb;8; and \verb;9; only finitely often. This is encoded as a parity objective with 3 priorities. State \verb'10' has priority 2, states \verb;8; and \verb;9; have priority 1, and every other state has priority 3. Whenever the goal state is reached, the maze is restarted 
with probability $1/3$ to state \verb;0;, with probability $1/3$ to state \verb;2;, and with probability $1/3$ to state \verb;4; in the easiest variant of the problem.

 The other variants are medium and difficult, and differ in the number of states the maze can restart to, intuitively increasing the uncertainty in the POMDP and also result into longer running times. We also consider a more difficult setting of the problem by constructing an intermediate and large size mazes with more states but based on the same principle. The input file for the POMDP can be downloaded here \url{http://pub.ist.ac.at/pps/examples/Small_cheese_maze_easy.txt}.

\subsection{Example - Grid}
We will describe only the grid $4\times4$, the other larger variants of the grid only differ in size. The problem consists of a $4$ by $4$ grid of locations. There is a single goal state, and multiple trap states that are placed beforehand but not known to the player. Whenever a goal state is reached the game is restarted. The goal of the player is to learn the position of the traps and visit the goal infinitely often, while visiting the trap state only finitely often with probability $1$.

\smallskip\noindent\textbf{States:} There are $33$ states in the POMDP. The uncertainty about the placement of the traps is modeled by two grids of states $4\times4$ and an additional initial state \verb;start; that has transition to both of the grids. The coding of the states is as follows, the state \verb;ijk; corresponds to a state in the $i$-th copy, $j$-th row, and $k$-th column. The goal state in each grid is the lower right corner, i.e., state \verb;033; and \verb;133; (rows and columns are numbered $0$-$3$).

\newcounter{row}
\newcounter{col}

\newcommand\setrow[4]{
  \setcounter{col}{1}
  \foreach \n in {#1, #2, #3, #4} {
    \edef\x{\value{col} - 0.5}
    \edef\y{3.5 - \value{row}}
    \node[anchor=center] at (\x, \y) {\n};
    \stepcounter{col}
  }
  \stepcounter{row}
}

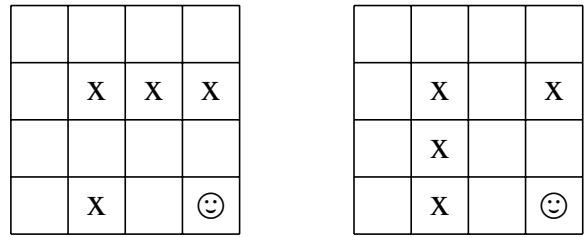
\begin{figure}[h]
\centering
\resizebox{8cm}{!}{
\begin{tikzpicture}[scale=.5]

  \begin{scope}
    \draw (0, 0) grid (4, 4);
    \setcounter{row}{0}
    \setrow {}{}{}{}
    \setrow {}{x}{x}{x}
    \setrow {}{}{}{}
    \setrow {}{x}{}{\smiley}




  \end{scope}
  \begin{scope}[xshift=6cm]
      \draw (0, 0) grid (4, 4);
    \setcounter{row}{0}
    \setrow {}{}{}{}
    \setrow {}{x}{}{x}
    \setrow {}{x}{}{}
    \setrow {}{x}{}{\smiley}
\end{scope}
\end{tikzpicture}
}
\caption{Trap placement}
  \label{fig:grid}

\end{figure}

\smallskip\noindent\textbf{Observations:}
There are $4$ observations, the initial state has observation \verb;o0;, the trap states have observation \verb;o2;, the goal state has observation \verb;o3;, and all the remaining states have observation \verb;o1;.

\smallskip\noindent\textbf{Actions:}
As in the previous example, there are $4$ actions available corresponding to the movement in the four compass directions (north \verb;n;, east \verb;e;, south \verb;s;, west \verb;w;).

\smallskip\noindent\textbf{Transition relation:}
The initial state of the POMDP is the state \verb;start;, no matter what action is played the next state is with probability $0.5$ the upper left corner in one of the grids, and with the remaining probability the upper left corner of the second grid. In the grid all
 the actions are deterministic and
attempts to move outside of the grid have no effect on the position. Whenever a goal state is reached the game is restarted the the upper left corner
of the same grid (the trap states stay in the same position).

\smallskip\noindent\textbf{Objective:} The objective is to learn in which grids the player is and as in the previous examples try to reach the goal state infinitely often while visiting the trap states only finitely often with probability $1$. This is encoded as a parity objective with $3$ priorities, the goal state has priority $2$, the trap states have priority $1$, and all the remaining states have priority $3$.

The placement of the trap states we have considered for the $4\times4$ grid is depicted on Figure~\ref{fig:grid}. The other variants differ in the size of the grid and placements of the trap states. The input file for the POMDP can be downloaded here \url{http://pub.ist.ac.at/pps/examples/4x4_grid.txt}.

\subsection{RockSample problems.}
We consider a modification of the RockSample problem introduced in~\cite{SS04} and used later in~\cite{BG09}. It is a scalable problem that models
 rover science exploration. The rover is equipped with a limited amount of fuel and can
 increase the amount of fuel by sampling rocks in the immediate area.
   The positions of the rover and the rocks are known, but only some of the rocks can increase the amount of fuel; we will call these 
   rocks good. The type of the rock is not known to the rover, until the rock is sampled.
Once a good rock is used to increase the amount of fuel, it becomes temporarily a bad rock until all other good rocks are sampled. We consider
   variants with different maximum capacity of the rover's fuel tank.
An instance of the RockSample problem is parametrized with two parameters $[n,k]$: map size $n \times n$ and $k$ rocks 
is described as RockSample[n,k]. The POMDP model of RockSample[n,k] is as follows:

\smallskip\noindent\textbf{States:}
The state space is the cross product of $2k+1+c$ features: 
$\pos = \{(1, 1), (1, 2), . . . , (n, n)\}$, $2*k$ binary features $\rockType_i = \{\goodObs, \badObs\}$ that indicate which of the rocks are good and which rocks
are temporarily not able to increase the amount of fuel, and $c$ is the amount of fuel remaining in the fuel tank.

\smallskip\noindent\textbf{Observations:} There are four observations: the unique observation for the initial state, two observations
to denote whether the rock that is sampled is good or bad. The last observation is for all the remaining states.

\smallskip\noindent\textbf{Actions:} The rover can select four actions: $\{N, S, E, W\}$.

\smallskip\noindent\textbf{Transition relation:} All the actions are deterministic single-step motion actions. 
A rock is sampled whenever the rock is at the rover's current location.

\smallskip\noindent\textbf{Objective:}
 If the rock is good, the fuel amount is increased to the maximum capacity and the rock becomes temporarily bad.
 Every state of the POMDP has priority $2$ with the following two exceptions:
\begin{itemize}
\item In a state where the rover samples a bad rock (also temporarily bad rocks) the priority $1$.
\item In a state where the fuel amount decreases to $0$ the priority is also 1.
\end{itemize}

The instance RS[4,2] (resp. RS[4,3]) is depicted on Figure~\ref{fig:rs_small} (resp. Figure~\ref{fig:rs_large}), the arrow indicates 
the initial position of the rover and the filled rectangles denote the fixed positions of the rocks.
A variant of the input file for the POMDP can be downloaded here \url{http://pub.ist.ac.at/pps/examples/RS4_2_3.txt}.

\begin{figure}[ht]
\centering
  \begin{minipage}[b]{0.40\linewidth} 
    \centering
    \resizebox{\linewidth}{!}{
    \begin{tikzpicture}[]
        \draw (0, 0) grid (4, 4);
        
        \fill [pattern=north west lines, pattern color=blue] (1,2) rectangle (2,3);
        \fill [pattern=north west lines, pattern color=blue] (3,0) rectangle (4,1);
        \draw[->]
        (-0.5,2.5) edge[ultra thick] node[above] {} (0.5,2.5);
        
        \draw [ultra thick] (0,0) rectangle (4,4);
    \end{tikzpicture}
    }
   \caption{RS[4,2]}
   \label{fig:rs_small}
  \end{minipage}
  \hspace{1.5cm}
  \begin{minipage}[b]{0.40\linewidth}
    \centering
 \centering
    \resizebox{\linewidth}{!}{
    \begin{tikzpicture}[]
        \draw (0, 0) grid (4, 4);
        \fill [pattern=north west lines, pattern color=blue] (1,3) rectangle (2,4);
        \fill [pattern=north west lines, pattern color=blue] (0,0) rectangle (1,1);
        \fill [pattern=north west lines, pattern color=blue] (2,1) rectangle (3,2);
        \draw[->]
        (-0.5,2.5) edge[ultra thick] node[above] {} (0.5,2.5);

        \draw [ultra thick] (0,0) rectangle (4,4);
    \end{tikzpicture}
    }
   \caption{RS[4,3]}
   \label{fig:rs_large}
  \end{minipage}
\end{figure}

\subsection{Hallway problems.}
We consider two versions of the Hallway problems introduced in~\cite{LCK95} and used later in~\cite{S04,SS04,BG09}.
The basic idea behind both of the Hallway problems, is that there is an agent 
wandering around an office building. 
It is assumed that the locations have been discretized so there are a finite number of 
locations where the agent could be. The agent has a small finite set of actions it can take, 
but these only succeed with some probability.  In these problems the location in the building and the agent's current orientation comprise the states.
The smaller Hallway POMDP is depicted in Figure~\ref{fig:hallway1} and the larger Hallway POMDP is depicted
in Figure~\ref{fig:hallway2}. We will describe the smaller problem Hallway in more detail:

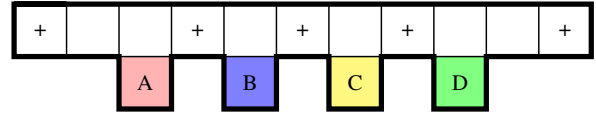
\begin{figure}[h]
\centering
\resizebox{8cm}{!}{

\begin{tikzpicture}
  \tikzstyle{every node}=[font=\large]
  \draw [fill=red, opacity=0.3] (2,1) rectangle (3,0);
  \draw [fill=blue, opacity=0.5] (4,1) rectangle (5,0);
  \draw [fill=yellow, opacity=0.5] (6,1) rectangle (7,0);
  \draw [fill=green, opacity=0.5] (8,1) rectangle (9,0);
  
  \draw[line width=0.1cm] (0,2)--(11,2)--(11,1)--(9,1)--(9,0)--(8,0)--(8,1)
  --(7,1)--(7,0)--(6,0)--(6,1)--(5,1)--(5,0)--(4,0)--(4,1)--(3,1)--(3,0)--(2,0)
  --(2,1)--(0,1)--(0,2)--(1,2);

  \draw[thick] (1,2)--(1,1);
  \draw[thick] (2,2)--(2,1);
  \draw[thick] (3,2)--(3,1);
  \draw[thick] (4,2)--(4,1);
  \draw[thick] (5,2)--(5,1);
  \draw[thick] (6,2)--(6,1);
  \draw[thick] (7,2)--(7,1);
  \draw[thick] (8,2)--(8,1);
  \draw[thick] (9,2)--(9,1);
  \draw[thick] (10,2)--(10,1);
  \draw[thick] (2,1)--(3,1);
  \draw[thick] (4,1)--(5,1);
  \draw[thick] (6,1)--(7,1);
  \draw[thick] (8,1)--(9,1);
  
  \node at (+2.5,+0.5) {A};
  \node at (+4.5,+0.5) {B};
  \node at (+6.5,+0.5) {C};
  \node at (+8.5,+0.5) {D};

  \node at (+0.5,+1.5) {+};
  \node at (+3.5,+1.5) {+};
  \node at (+5.5,+1.5) {+};
  \node at (+7.5,+1.5) {+};
  \node at (+10.5,+1.5) {+};

\end{tikzpicture}
}
\caption{Hallway POMDP}
\label{fig:app_hallway1}
\end{figure}

\smallskip\noindent\textbf{States:}
There are $15$ locations in the office times the four possible orientations together with an auxiliary starting and loosing state is $62$. All the
objectives are expressed as deterministic parity automata. The final size of the POMDP $62$ multiplied by the number of states
of the parity automaton.

\smallskip\noindent\textbf{Observations:} the agent is equipped 
with very short range sensors to provide it only with information about whether it is 
adjacent to a wall. The sensors can ''see'' in four directions: forward, left, right,  and backward.
 It is important to note that these observations are relative to the 
current orientation of the agent (N, E, S, W).

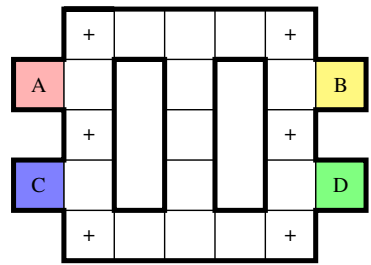
\begin{figure}[h]
\centering
\resizebox{5cm}{!}{

\begin{tikzpicture}
  \tikzstyle{every node}=[font=\large]
  \draw [fill=red, opacity=0.3] (1,5) rectangle (2,4);
  \draw [fill=blue, opacity=0.5] (1,3) rectangle (2,2);
  \draw [fill=yellow, opacity=0.5] (7,5) rectangle (8,4);
  \draw [fill=green, opacity=0.5] (7,3) rectangle (8,2);
  
  \draw[line width=0.1cm] (2,6)--(7,6)--(7,5)--(8,5)--(8,4)--(7,4)--(7,3)--(8,3)--(8,2)
  --(7,2)--(7,1)--(2,1)--(2,2)--(1,2)--(1,3)--(2,3)--(2,4)--(1,4)--(1,5)--(2,5)--(2,6)--(3,6);
  \draw[line width=0.1cm] (3,5)--(4,5)--(4,2)--(3,2)--(3,5)--(4,5);
  \draw[line width=0.1cm] (5,5)--(6,5)--(6,2)--(5,2)--(5,5)--(6,5);
  \draw[thick] (2,5)--(3,5);
  \draw[thick] (4,5)--(5,5);
  \draw[thick] (6,5)--(7,5);
  \draw[thick] (2,4)--(3,4);
  \draw[thick] (4,4)--(5,4);
  \draw[thick] (6,4)--(7,4);
  \draw[thick] (2,3)--(3,3);
  \draw[thick] (4,3)--(5,3);
  \draw[thick] (6,3)--(7,3);
  \draw[thick] (2,2)--(3,2);
  \draw[thick] (4,2)--(5,2);
  \draw[thick] (6,2)--(7,2);
  \draw[thick] (2,5)--(2,4);
  \draw[thick] (2,3)--(2,2);
  \draw[thick] (7,5)--(7,4);
  \draw[thick] (7,3)--(7,2);
  \draw[thick] (3,6)--(3,5);
  \draw[thick] (4,6)--(4,5);
  \draw[thick] (5,6)--(5,5);
  \draw[thick] (6,6)--(6,5);
  \draw[thick] (3,2)--(3,1);
  \draw[thick] (4,2)--(4,1);
  \draw[thick] (5,2)--(5,1);
  \draw[thick] (6,2)--(6,1);
  
  \node at (+1.5,+4.5) {A};
  \node at (+7.5,+4.5) {B};
  \node at (+1.5,+2.5) {C};
  \node at (+7.5,+2.5) {D};

  \node at (+2.5,+5.5) {+};
  \node at (+6.5,+5.5) {+};
  \node at (+2.5,+3.5) {+};
  \node at (+6.5,+3.5) {+};
  \node at (+2.5,+1.5) {+};
  \node at (+6.5,+1.5) {+};

\end{tikzpicture}
}
\caption{Hallway 2 POMDP}
\label{fig:app_hallway2}
\end{figure}

\smallskip\noindent\textbf{Actions:}
There are three actions that can be chosen: forward, turn-left, and turn-right.

\smallskip\noindent\textbf{Transition relation:}
The agent starts with uniform probability in the states labeled with the + symbol in any of the four possible orientations.
The actions that can be chosen consists of movements: forward, turn-left, and turn-right.
All the available actions succeed with probability $0.8$  and with probability $0.2$ 
the state is not changed. In states where moving forward is impossible the probability mass 
for the impossible next state is collapsed into the probability of not changing the state.

\smallskip\noindent\textbf{Objective:}
There are four dedicated areas in the office, denoted by letters A,B,C, and D. We consider four objectives in both the 
Hallway problems:
\begin{itemize}
\item \emph{Liveness:} requires that the $D$-labeled state is reached. The automaton consists of $2$ states.
\item \emph{Sequencing and avoiding obstacles:} requires that first the $A$-labeled state is visited, followed by the $B$-labeled state
and finally the $D$-labeled state is visited while avoiding the $C$-labeled state. The automaton consists of $5$ states.
\item \emph{Coverage:} requires that the $A$, $B$, and $C$-labeled states are all visited in any order. The automaton consists of $8$ states.
\item \emph{Recurrence:} requires that both the $A$ and $C$-labeled states are visited infinitely often. The automaton consists of $4$ states.
\item \emph{Recurrence and avoidance:} requires that both $A$ and $D$-labeled states are visited infinitely often, while visiting $B$ and $C$-labeled states
only finitely many times. The automaton consists of $5$ states.
\end{itemize}
A variant of the input file of the POMDP can be downloaded here \url{http://pub.ist.ac.at/pps/examples/hallwayLiv.txt}.

\subsection{Maze navigation problems.}
We consider three variants of the mazes introduced in~\cite{GMK13}. Intuitively, the robot navigates itself in a grid
discretization of a 2D world. The robot can choose from four noise free actions north, east, south, and west. In every maze
there are $4$ highlighted regions that are labeled with letters $A$, $B$, $C$, and $D$. The objective for the robot is given
as a deterministic parity automaton, as in the case of Hallway problems.
We describe fully Maze~A, the other problems differ only in the structure of the maze:

\begin{figure}[ht]
\centering
  \begin{minipage}[b]{0.40\linewidth} 
    \centering
    \resizebox{\linewidth}{!}{

\begin{tikzpicture}
  \tikzstyle{every node}=[font=\large]
  \draw [fill=red, opacity=0.3] (0,8) rectangle (2,9);
  \draw [fill=blue, opacity=0.5] (8,8) rectangle (10,9);
  \draw [fill=yellow, opacity=0.5] (3,4) rectangle (8,5);
  \draw [fill=green, opacity=0.5] (0,0) rectangle (10,1);
  \draw [fill=black] (2,9) rectangle (3,3);
  \draw [fill=black] (3,3) rectangle (7,4);
  \draw [fill=black] (7,3) rectangle (8,9);
  \draw [thick] (0,0) grid (10,10);
  \node at (+4.5,+6.5) {+};
  \node at (+6.5,+6.5) {+};
  \node at (+4.5,+8.5) {+};
  \node at (+6.5,+8.5) {+};
  \node at (+0.5,+8.5) {A};
  \node at (+1.5,+8.5) {A};
  \node at (+8.5,+8.5) {B};
  \node at (+9.5,+8.5) {B};
  \node at (+3.5,+4.5) {C};
  \node at (+4.5,+4.5) {C};
  \node at (+5.5,+4.5) {C};
  \node at (+6.5,+4.5) {C};
  \node at (+0.5,+0.5) {D};
  \node at (+1.5,+0.5) {D};
  \node at (+2.5,+0.5) {D};
  \node at (+3.5,+0.5) {D};
  \node at (+4.5,+0.5) {D};
  \node at (+5.5,+0.5) {D};
  \node at (+6.5,+0.5) {D};
  \node at (+7.5,+0.5) {D};
  \node at (+8.5,+0.5) {D};
  \node at (+9.5,+0.5) {D};
\end{tikzpicture}
    }
    \caption{POMDP Maze A}
	\label{fig:mazeA}

  \end{minipage}
  \hspace{1.5cm}
  \begin{minipage}[b]{0.40\linewidth}
    \centering
 \centering
    \resizebox{\linewidth}{!}{

\begin{tikzpicture}
  \tikzstyle{every node}=[font=\large]
  \draw [fill=red, opacity=0.3] (8,2) rectangle (10,3);
  \draw [fill=blue, opacity=0.5] (4,9) rectangle (6,10);
  \draw [fill=yellow, opacity=0.5] (3,5) rectangle (7,6);
  \draw [fill=green, opacity=0.5] (0,0) rectangle (10,1);
  \draw [fill=black] (0,1) rectangle (4,3);
  \draw [fill=black] (7,1) rectangle (10,2);
  \draw [fill=black] (7,1) rectangle (8,7);
  \draw [fill=black] (3,6) rectangle (7,7);
  \draw [thick] (0,0) grid (10,10);
  \node at (+8.5,+2.5) {A};
  \node at (+9.5,+2.5) {A};
  \node at (+8.5,+3.5) {+};
  \node at (+9.5,+3.5) {+};

  \node at (+4.5,+9.5) {B};
  \node at (+5.5,+9.5) {B};
  \node at (+3.5,+5.5) {C};
  \node at (+4.5,+5.5) {C};
  \node at (+5.5,+5.5) {C};
  \node at (+6.5,+5.5) {C};
  \node at (+0.5,+0.5) {D};
  \node at (+1.5,+0.5) {D};
  \node at (+2.5,+0.5) {D};
  \node at (+3.5,+0.5) {D};
  \node at (+4.5,+0.5) {D};
  \node at (+5.5,+0.5) {D};
  \node at (+6.5,+0.5) {D};
  \node at (+7.5,+0.5) {D};
  \node at (+8.5,+0.5) {D};
  \node at (+9.5,+0.5) {D};
\end{tikzpicture}
    }
    \caption{POMDP Maze B}
	\label{fig:mazeB}
  \end{minipage}
\end{figure}

\smallskip\noindent\textbf{States:}
The state space of Maze~A consists of $84$ possible grid locations times the number of states of the parity automaton that defines
the objective. The robot moves from the unique initial states uniformly at random under all actions in all the locations labeled with "+".

\smallskip\noindent\textbf{Observations:}
The highlighted regions are observable to the robot, otherwise the robot does not receive any feedback from the maze.

\smallskip\noindent\textbf{Actions:}
There are four actions that can be chosen: $\{N, S, E, W\}$.

\smallskip\noindent\textbf{Transition relation:}
All the available actions succeed with probability $1$. In states where robot attempts to move outside of the maze or 
in the wall the position of the robot remains unchanged.

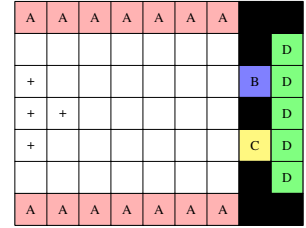
\begin{figure}[h]
\centering
\resizebox{4cm}{!}{

\begin{tikzpicture}
  \tikzstyle{every node}=[font=\large]
  \draw [fill=red, opacity=0.3] (0,0) rectangle (7,1);
  \draw [fill=red, opacity=0.3] (0,6) rectangle (7,7);
  \draw [fill=green, opacity=0.5] (8,1) rectangle (9,6);
  \draw [fill=black] (7,5) rectangle (8,7);
  \draw [fill=black] (8,6) rectangle (9,7);
  \draw [fill=black] (7,3) rectangle (8,4);
  \draw [fill=yellow, opacity=0.5] (7,2) rectangle (8,3);
  \draw [fill=blue, opacity=0.5] (7,4) rectangle (8,5);
  \draw [fill=black] (7,0) rectangle (8,2);
  \draw [fill=black] (8,0) rectangle (9,1);
  \draw [thick] (0,0) grid (9,7);
  \node at (+0.5,+0.5) {A};
  \node at (+1.5,+0.5) {A};
  \node at (+2.5,+0.5) {A};
  \node at (+3.5,+0.5) {A};
  \node at (+4.5,+0.5) {A};
  \node at (+5.5,+0.5) {A};
  \node at (+6.5,+0.5) {A};

  \node at (+0.5,+6.5) {A};
  \node at (+1.5,+6.5) {A};
  \node at (+2.5,+6.5) {A};
  \node at (+3.5,+6.5) {A};
  \node at (+4.5,+6.5) {A};
  \node at (+5.5,+6.5) {A};
  \node at (+6.5,+6.5) {A};
  \node at (+7.5,+4.5) {B};
  \node at (+7.5,+2.5) {C};

  \node at (+8.5,+1.5) {D};
  \node at (+8.5,+2.5) {D};
  \node at (+8.5,+3.5) {D};
  \node at (+8.5,+4.5) {D};
  \node at (+8.5,+5.5) {D};
  \node at (+0.5,+2.5) {+};
  \node at (+0.5,+3.5) {+};
  \node at (+0.5,+4.5) {+};
  \node at (+1.5,+3.5) {+};
\end{tikzpicture}
}
\caption{POMDP Maze C}
\label{fig:mazeC}
\end{figure}

\smallskip\noindent\textbf{Objective:}
We consider the same objectives as in the case of the Hallway problems:
\begin{itemize}
\item \emph{Liveness:} requires that the $D$-labeled state is reached. The automaton consists of $2$ states.
\item \emph{Sequencing and avoiding obstacles:} requires that first the $A$-labeled state is visited, followed by the $B$-labeled state
and finally the $D$-labeled state is visited while avoiding the $C$-labeled state. The automaton consists of $5$ states.
\item \emph{Coverage:} requires that the $A$, $B$, and $C$-labeled states are all visited in any order. The automaton consists of $8$ states.
\item \emph{Recurrence:} requires that both the $A$ and $C$-labeled states are visited infinitely often. The automaton consists of $4$ states.
\item \emph{Recurrence and avoidance:} requires that both $A$ and $D$-labeled states are visited infinitely often, while visiting $B$ and $C$-labeled states
only finitely many times. The automaton consists of $5$ states.
\end{itemize}
A variant of the input file of the POMDP can be downloaded here \url{http://pub.ist.ac.at/pps/examples/mazeALiv.txt}.

\end{document}